\documentclass[prc,twocolumn]{revtex4}
\usepackage{graphicx,colordvi}
\usepackage[dvips]{color}
\usepackage{amsmath}

\begin{document}
%                               *** a 2 Seiten NF901TIT einfuegen 
%\clearpage
%\thispagestyle{empty}
%\vspace*{0.5cm}

\title{Is simultaneous $y$ and $\xi$--scaling in the quasi-elastic region accidental?}
\author{Donal Day}\affiliation{Dept. of Physics, Univ. of Virginia, Charlottesville, VA, USA}
\author{Ingo Sick}
\affiliation{Dept. f\"ur Physik und Astronomie, Univ. Basel, Basel, Switzerland}
\begin{abstract}We study the $y$ and $\xi$--scaling of the 
nuclear response at large momentum transfer in order to understand how scaling based on very different descriptions of the  elementary interaction can occur simultaneously. We find that the approximate validity of $\xi$-scaling at
low energy loss  arises from the coincidental behavior of the quasielastic and deep inelastic cross sections.\end{abstract}
\maketitle

\section{Introduction}
Scaling occurs in several fields where a probe scatters incoherently from a 
composite system: quasi-elastic scattering of electrons from nucleons bound 
in nuclei \cite{Sick80}, deep inelastic scattering (DIS) of leptons from 
quarks bound in nucleons \cite{Bjorken69},
scattering of thermal neutrons from individual atoms in liquids such as L$^4$He
\cite{Sokol89,Sears84}, and Compton scattering from atoms \cite{Eisenberger70}, 
to name a few. These inclusive scattering cross sections {\em a priori} are 
functions of two independent variables, the momentum transfer $q$ and 
the energy transfer $\nu$. The scattering is said to scale if the cross sections, after removal (by division)
 of the elementary probe-constituent cross section and a kinematic factor
  %(\textcolor{red}{arising from the integration over the energy and momentum conserving delta function})
  determined by the elastic scattering kinematics, become
a function of a {\em single} variable $z = z(q,\nu)$ rather than $q$ and $\nu$
separately \cite{Day90}. 
%Scaling appears as a consequence of the removal of the momentum dependence of 
%the elementary probe-constituent cross section and the kinematics
%of probe-constituent elastic scattering relating energy- and momentum-transfer.
  The observation of scaling (and of scaling violations)  provides a better 
understanding 
of both the reaction mechanism and the structure of the system in which the constituent is bound.

Quasi-elastic electron-nucleus scattering shows a scaling behavior once the inclusive 
scattering cross sections are divided by the 
%(appropriate combination of)
elastic electron-nucleon cross section and a kinematic factor connecting the scaling variable and energy and momentum transfer. The 
corresponding scaling variable 
$y\equiv - q + (\nu^2 + 2 m \nu)^{1/2}   $
is obtained from energy/momentum conservation for elastic scattering from 
an initially off--shell nucleon, and the scaling function $F(y)$ represents the 
probability to find nucleons in the nucleus with the minimum momentum $y$ 
 allowed by the kinematics. The minimum momentum, $y$,  is  parallel to $\vec{q}$.
 
 Scaling of the nucleon structure functions, $M W_1$ and $\nu W_2$, in DIS is revealed after division of 
the inclusive cross section by the Mott cross section (describing the $Q^2$ dependence of  point-like
scattering). The scaling variable in DIS,
$x = Q^2/2 m \nu$,
($Q$ is the four-momentum transfer) is the struck parton's fraction of the nucleon momentum in the infinite momentum frame. The scaling function $\nu W_2$ allows access to the quark distribution functions. The Nachtmann variable \cite{Nachtmann73}  
$\xi = (q - \nu)/m$ (other equivalent expressions are customary 
in the literature)
is often used to extend DIS scaling to lower momentum transfers.

In the kinematical regions dominated by quasi-elastic  scattering and DIS,
 scaling is observed in $y$ and $x$ (or $\xi$), respectively. Deviations from ideal scaling in
quasi-elastic scattering are due to the effects of the final-state interaction 
(FSI) of the recoiling nucleon, and the importance of large removal energies
$E$ in the nuclear spectral function $S(k,E)$ \cite{Day90}. For DIS, scaling violations mainly
arise from the evolution of the structure functions with $Q^2$   and are well 
understood within perturbative QCD
\cite{Ellis96}.
When neglecting FSI,  $\nu W_2$ provides information on the quark 
distribution functions \cite{Brodsky02}.   $\xi$, for
a mass-zero constituent initially off-shell, is its momentum component  
parallel to $\vec{q}$, in units of the nucleon mass m.

{\bf $\xi$--scaling}
Recent work on $\xi$-scaling has been focused on exploring both  resonance region and the DIS region. It has 
been found that the response averaged over the resonances defines the same scaling 
function as derived from DIS at very large momentum transfer and evolved  to
 $Q^2$ appropriate for the resonances \cite{Niculescu00a,Niculescu00b}. 
This observation, first revealed observed some years ago, and referred to as 
``duality'' \cite{Bloom71}, is only partly understood. The extension of such studies to 
very low energy loss, the region of the quasi-elastic peak in nuclei at $y<0$ or $x>1$,
showed \cite{Filippone92} that $\xi$-scaling apparently works even where the dynamics of 
probe-constituent scattering (quasi-elastic scattering from a nucleon) is very different from the one used in $\xi$-scaling.

The observation of $\xi$-scaling in the region of the low-$\nu$ side of the
quasi-elastic peak certainly is puzzling. $y$-scaling in this region is dependent
on the use of elastic electron-nucleon kinematics for the scattering process, which 
is quite different for DIS as the former involves a dependence on the finite nucleon mass
(at the $q$'s of interest here). As importantly, the elementary 
probe-constituent cross section momentum dependences are quite dissimilar. In  $\xi$-scaling the Mott cross section carries all the $Q^2$ dependence while for $y$-scaling the electron--nucleon cross sections depend  on the nucleon elastic form factors which carry an additional strong $Q^2$ dependence.

Benhar and Liuti \cite{Benhar95c} already have noted this puzzle and suggested an explanation. For quasi-elastic scattering at moderately large $q$ the FSI of 
the recoiling nucleon still plays an important role \cite{Benhar91}. 
This FSI leads    
to a violation of $y$-scaling. Benhar and Liuti showed that, for the inclusive cross sections
then available, the $Q^2$ dependences of the FSI and of  the relationship between $\xi$ and $y$ could combine such that there would be, accidentally,    an approximate scaling in terms of $\xi$.
% The FSI 
%contribution could lead to an unexpected scaling when
%studying quasi-elastic scattering from nuclei in terms of the DIS scaling 
%variable $\xi$ derived for massless constituents.

An experiment (E89-008) performed at Jefferson Lab provided a much a more extensive set of data\cite{Arrington99}
on quasi-elastic electron-nucleus scattering. This data extends to much larger
momentum transfers and, for a given $y$ or $\xi$, covers a larger range in $Q^2$. 
These data revealed, as expected, $y$-scaling on the low-$\nu$ side of the quasi-elastic
peak. Surprisingly it also displayed an approximate scaling in terms of $\xi$
which, at the largest $q$, extends to rather low values of $\nu$. The authors of 
\cite{Arrington99} pointed out that over this larger range of $q$ the 
explanation for
scaling in terms of in $\xi$ provided by Benhar and Liuti could no longer be valid; at large $Q^2$ the FSI
decreases, while the elementary cross sections and the kinematics describing the reaction for $\xi$ and $y$ scaling remain disparate. The simultaneous scaling in $y$ and $\xi$ thus remains a puzzle, and 
it is not clear whether $\xi$ scaling in the quasi-elastic region occurs due to some accidental convergence of factors or 
whether it has a deeper meaning.

In this Letter, we return to this question of ``accidental'' $\xi$-scaling to
provide additional insight. 

\section{Pieces of the Puzzle} 
\begin{figure}[htb] 
\begin{center}
\includegraphics[width=8cm,clip]{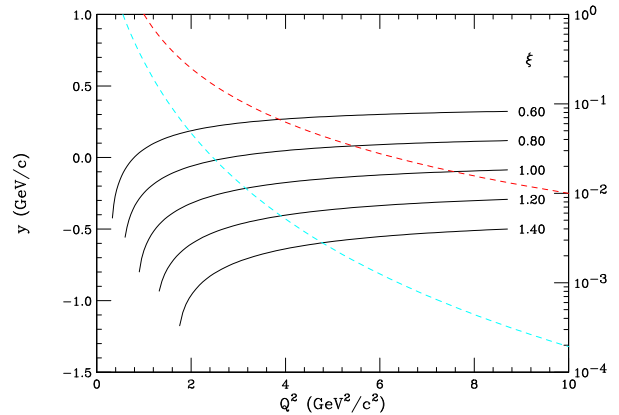}
\caption{Relation between $\xi$ and $y$ as a function of momentum transfer 
$Q^2$. The dotted curves (right hand scale) represent the $Q^2$-dependence of 
the Mott cross
section (upper curve) and the nucleon form factor squared.}\label{fig:xiys}
\end{center}
\end{figure}

% A BETTER TITLE HERE WOULD BE WELCOME
%
An understanding of the simultaneous display of $y$  and $\xi$ scaling in the quasi-elastic region requires going beyond the question that
was addressed by Benhar and Liuti \cite{Benhar95c}. The definitions of  $y$ and $\xi$ provide part of the understanding as they lead to a relative shift of data points
taken at different $Q^2$'s, but that can not explain the phenomenon. The $y$-scaling function, $F(y)$
is obtained by dividing the inclusive cross section by the  cross section for incoherent 
electron-nucleon elastic scattering (summed over the number of protons and neutrons).  In the case of the $\xi$-scaling function, $\nu W_2$, only the Mott cross section 
is divided out. These two cross sections differ by the nucleon elastic 
form factor squared, approximately proportional to
$1./(1.+ Q^2/0.71)^4$. (See Fig.~\ref{fig:xiys}.)
This factor is responsible for an additional $Q^2$-dependence in the case of 
quasi-elastic scattering and is not included in a $\xi$-scaling analysis. Thus $\nu W_2^A$ should show a pronounced scaling violation in the quasi-elastic region. 

In the limit of extremely large $Q^2$ (once the 
momentum transfer is much larger than the nucleon mass) the variable  $y$ converges toward 
$(-\xi + 1) m$. In this  case $y$ and 
$\xi$ would become equivalent. Significantly, the difference in the elementary cross sections  does remain 
at the $q$'s of interest and should continue to forbid simultaneous scaling in terms
of $y$ and $\xi$.

{\bf Relation between $\xi$ and $y$~~}
In order to better understand the difference between $y$ and $\xi$ scaling, 
  we first discuss the relation between the two variables. Scaling in terms of $\xi$ 
involves a study of the nuclear response at $\xi$=const, as a function of the
momentum transfer.  Fig.~\ref{fig:xiys} shows the relation between $\xi$ and $y$ and it can be seen that for a given value of $\xi$,  $y$ is a pronounced function  of $Q^2$, particularly at the lower momentum  transfers. 

As a consequence, the nuclear DIS scaling function, $\nu W_2^A$, at constant value of $\xi$, includes contributions from a large range of $y$ values when studied over  $Q^2$. The 
nuclear quasi-elastic scaling function $F(y)$ (roughly speaking the probability to find in the 
nucleus nucleons with momentum component $y$ parallel to $q$) depends strongly on $y$, as nuclear momentum distributions are a steeply falling function 
with increasing $|y|$.  How $\xi$-scaling appears in the quasi-elastic region requires an understanding of the behavior of several rapidly varying functions --  $\sigma_{eN}(Q^2)$, $\sigma_{Mott}(Q^2)$  and $F(y)$.

\section{Model studies} 
In order to understand expose the behavior of the different scaling variables and scaling 
functions, we made a study of the nuclear response using a simple model. In this
model, we describe the inclusive nuclear response as a sum of quasi-elastic
scattering, excitation of the nucleon resonances and DIS on the nucleon.

For quasi-elastic scattering our model is based on $y-$scaling at intermediate 
momentum transfers.  $F(y)$  is determined from the experimental data through  
$$F(y)=\frac{d^2 \sigma^{\textrm{exp}}}{d \Omega dE'}/ \sigma_{ei} \cdot K  $$ 
where $\sigma_{ei}$ is the appropriate combination of neutron and proton
elastic cross sections and 
$K =|{\bf q}| / \sqrt{m^2 + (y + |{\bf q}|)^2}$
is the Jacobian resulting from the change of variables. We fit $F(y)$  with the sum of a Gaussian and an exponential 
function and limit the fit essentially to the
$y<0$ region. As long as we can neglect the effect of the nucleon removal energy $E$ 
(beyond the average $\bar{E}$ which we do take into account) the response
is symmetric about $y=0$. The resulting  momentum distribution, $n(k) = -\frac{1}{2\pi y}\frac{dF(y)}{dy}$
is a good approximation of
what is expected  from theoretical calculations of $n(k)$.  

For the resonance and DIS regions the neutron and proton responses have been 
parameterized by Bodek and Ritchie \cite{Bodek81} by fitting inclusive 
scattering data on the nucleon 
and removing, for the case of the neutron, the Fermi motion effects occurring
in the deuteron structure function. To calculate the nuclear response,
we fold the nucleonic response of Ref.~\cite{Bodek81} with the nuclear momentum distribution $n(k)$ determined above.

The sum of these contribution reproduces the experimental nuclear responses
quite well. Typical deviations between model and data are of the order of 20\%.

\begin{figure}[htb] 
\begin{center}
\includegraphics[angle=90,width=8cm,clip]{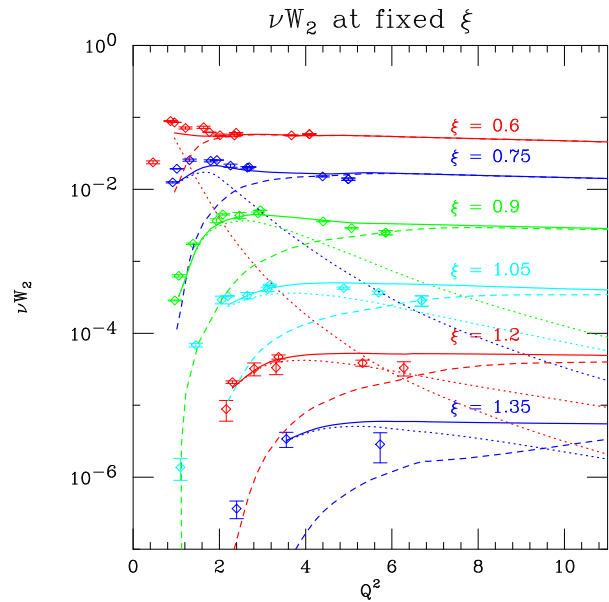}
\caption{$\nu W_2$ at fixed values of $\xi$ as a function of $Q^2$ for iron. Data from 
SLAC NE3 and Jefferson Lab E89-008 are plotted along with the results of the 
model described in the text. Dotted lines reflect the quasi-elastic piece, 
dashed lines the inelastic and the solid line the sum.\label{fig:fe-xsi}
The data sets correspond to $\xi= 0.6, 0.75, 0.9, 1.05, 1.2   \textrm{ and } 1.35$. }\end{center}
\end{figure}

In what follows below we determine $\nu W_2^A$ from the inclusive cross section $\sigma$ (both model and experimental) via 
$$\nu W_2^A(Q^2)= \nu \cdot \frac{\sigma}{\sigma_M} \biggl[ 1 + 2 \tan^2(\theta/2) \cdot \left( \frac{1+\nu^2/Q^2}{1+R} \right )  \biggr]^{-1} $$
where $\sigma_{M}$ is the cross section for point like constituents and  the ratio of longitudinal to transverse cross section, $R= \sigma_L/\sigma_T = (1+\nu^2/Q^2)W_2/W_1 -1$. We take $R= \frac{0.32}{Q^2}$. The model cross sections are the sum of the quasi-elastic and inelastic contributions as described above.
We display in Fig.~\ref{fig:fe-xsi} the model scaling function $\nu W_2$ for iron (per nucleon) for six different $\xi$ values as a function of the momentum transfer along with data from Jefferson Lab 
E89-008 \cite{Arrington99} and SLAC 
NE3 \cite{Day87,Day93}. At very
large $Q^2$ $\xi = 1$ corresponds to the top of the quasi-elastic
peak ($y=0$, $x=1$) and at  lower momentum transfers $\xi = 1$ maps 
to increasingly more negative values of $y$ (see Fig.~\ref{fig:xiys}).

From Fig.~\ref{fig:fe-xsi}  we see that the sum of the quasi-elastic 
response and the combined resonance and DIS contributions scales better than the
 combined resonance and DIS piece alone.    The latter falls steeply at low $Q^2$. 
This fall-off is, to a large degree, just compensated by the quasi-elastic 
contribution. The reduced $Q^2$-dependence of the sum leads to the approximate 
$\xi$-scaling observed experimentally.

At fixed $\xi$, increasing $Q^2$ shifts $y$ 
to less negative values, where $F(y)$ is larger. At the same
time, this increase in the $F(y)$ is largely offset (for quasielastic scattering) by the nucleon elastic form factors
which fall with increasing $Q^2$.  The DIS contribution to $\nu W_2^A$ at fixed $\xi$ grows with $F(y)$ ($n(k)$)  as $Q^2$ increases but does not suffer from the $Q^2$ dependence of the elastic form factors. Taken together, these two incongruous behaviors lead to an approximately constant $\nu W_2^A$. 
 
While the shift of $y$ with $Q^2$ is of purely kinematical origin, the decrease of 
$F(y)$ with larger $|y|$ (the decrease of $n(k)$ with larger $k$) depends 
entirely on the behavior of the nuclear momentum distribution. The 
magnitude of this fall-off is specific to the nucleus under consideration, although for different
 nuclei the
tails of nuclear momentum distributions at large $k$ are not too different given
the identical origin of these tails, the short-range nucleon-nucleon correlations. 
There is, however, no connection we can discern that relates the physics 
responsible for the fall-off of $n(k)$ to the $Q^2$-dependence of the nucleon form 
factors and the kinematics that distinguishes $\xi$ from $y$. From this
observation we conclude that the apparent $\xi$-scaling is simply due to how the quasi-elastic piece and resonance and DIS pieces sum together at fixed $\xi$  and is largely accidental in character.

\begin{figure}[htb] 
\begin{center}
\includegraphics[angle=90,width=8cm,clip]{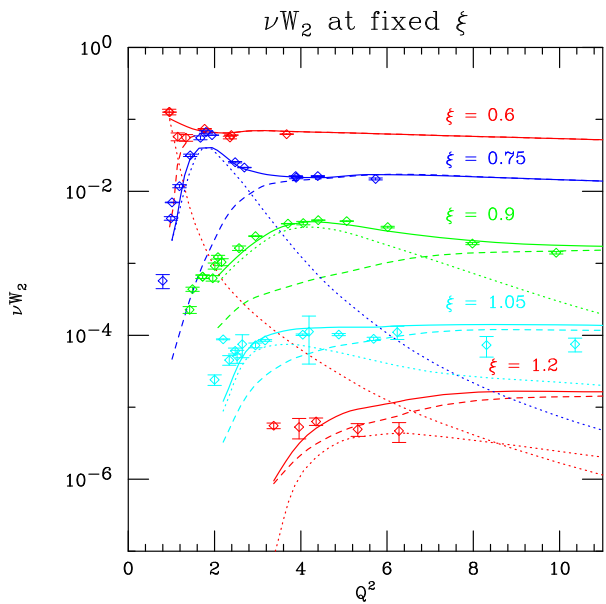}
\caption{\label{fig:deut-xsi}$\nu W_2$ for the deuteron at fixed values of 
$\xi$ as a function of $Q^2$. Data from SLAC experiments E101, E133, and NE4 
and Jefferson Lab E89-008 are plotted along with the results of the  model 
described in the text. Dotted lines reflect the quasi-elastic piece, dashed 
lines the inelastic and the solid line the sum.}
\end{center}
\end{figure}

This point on the dependence on $F(y)$ ($n(k)$)  is further emphasized by 
Fig.~\ref{fig:deut-xsi} which shows our model $\nu W_2$ for the deuteron 
along with data from 
\cite{Schuetz77,Arnold88,Rock92,Arrington99}.
In the corresponding range of $y$ or $k\ (0.3 < k < 0\ \mathrm{ GeV/c)}$ the deuteron
momentum distribution falls off more quickly with increasing $y$ (or $k$) than in heavier nuclei. Accordingly,
the approach to $\xi$-scaling is different: 
%\textcolor{red}{
the rapid fall-off of $n(k)$ 
actually better compensates the fall-off of the nucleon form factor with
$Q^2$, and leads  to significantly better scaling in terms of $\xi$.
%}

\section{Conclusions}
We have studied the experimental observation of $\xi$-scaling in 
electron-nucleus scattering in the region of $q, \nu$ where the cross 
section would not be expected to   scale as it is dominated quasi-elastic electron-nucleon scattering. We find 
that the apparent $\xi$-scaling comes about because the quasi-elastic 
contribution approximately compensates for the failure of resonance and DIS contribution at large $\xi$ and low $Q^2$
to scale. 
The two dominant, rapidly varying functions distinguishing the behavior of the
nuclear scaling function, $\nu W_2$, as a function of $y$ and $\xi$ are the nuclear momentum distribution  
 at low $k$  and the nucleon form factors at large $Q^2$. As these have no obvious physical connection, we conclude that the 
appearance of an approximate $\xi$-scaling at large $\xi$ comes about accidentally--the two dominant contributions to the inclusive cross section behave such that their sum shows a $Q^2$ independence  characteristic of scaling, but separately they do not. 

It remains to be seen whether the concept of duality --- which
refers to an {\em average} over many resonances rather than a single one (the 
nucleon ground state) --- can further elucidate this accidental $\xi$-scaling.

\begin{acknowledgements} This work was supported by the U.S. Department of 
Energy and the Schweizerische Nationalfonds.\end{acknowledgements} 
%\input{out.bbl}
%\bibliographystyle{unsrt}
%\bibliography{/usr/users/sick/sum2}

\begin{thebibliography}{99}                                                     
\bibitem{Sick80}                                                                
I.~Sick {\em et al.}, Phys. Rev. Lett., 45:871, 1980.                                 

\bibitem{Bjorken69}                                                             
J.D. Bjorken and E.A. Paschos, Phys. Rev., 185:1975, 1969.                                     

\bibitem{Sokol89}                                                               
P.E. Sokol {\em et al.},  Momentum Distributions, Plenum Press, Edts R. Silver, P.E. Sokol         

\bibitem{Sears84}                                                               
V.F. Sears,  Physical Review B, 30:44, 1984.                                 


\bibitem{Eisenberger70}                                                         
P.~Eisenberger and P.M. Platzman, Phys. Rev. A, 2:415, 1970.                                      


\bibitem{Day90}                                                                 
D.~Day {\em et al.},  Ann. Rev. Nucl. Part. Sci., 40:357, 1990.                       


\bibitem{Nachtmann73}                                                           
O.~Nachtmann,  Nucl. Phys. B, 63:237, 1973.                                    


\bibitem{Ellis96}                                                               
R.~K. Ellis {\em et al.},  QCD and Collider Physics, Cambridge Univ. Press, 1996.          


\bibitem{Brodsky02}                                                             
S.J.Brodsky {\em et al.}, Phys. Rev. D, 65:114025, 2002.                                  


\bibitem{Niculescu00a}                                                          
I.~Niculescu {\em et al.}, Phys. Rev. Lett., 85:1182, 2000.                               


\bibitem{Niculescu00b}                                                          
I.~Niculescu {\em et al.}, Phys. Rev. Lett., 85:1186, 2000.                                


\bibitem{Bloom71}                                                               
E.D.Bloom and F.J.Gilman,  Phys. Rev. D, 4:2901, 1971.                                     


\bibitem{Filippone92}                                                           
B.W. Filippone {\em et al.}, Phys. Rev. C, 45:1582, 1992.                                    


\bibitem{Benhar95c}                                                             
O.~Benhar and S.~Liuti, Phys. Lett. B, 358:173, 1995.                                   


\bibitem{Benhar91}                                                              
O.~Benhar {\em et al.}, Phys. Rev. C, 44:2328, 1991.                                    


\bibitem{Arrington99}                                                           
J.~Arrington {\em et al.}, Phys. Rev. Lett., 82:2056, 1999.                               


%\bibitem{Benhar95d}                                                             
%O.~Benhar and S.~Liuti.                                                         
%\newblock {\em Phys. Lett. B}, 358:173, 1995.                                   


\bibitem{Bodek81}                                                               
A.~Bodek and J.L. Ritchie, Phys. Rev. D, 23:1070, 1981.                                    


\bibitem{Day87}                                                                 
D.~Day {\em et al.}, Phys. Rev. Lett., 59:427, 1987.                                 


\bibitem{Day93}                                                                 
D.~Day {\em et al.}, Phys. Rev. C, 48:1849, 1993.                                    


\bibitem{Schuetz77}                                                             
W.P.~Schuetz {\em et al.}, Phys. Rev. Lett., 38:259, 1977.                                 


\bibitem{Arnold88}                                                              
R.G. Arnold {\em et al.},  Phys. Rev. Lett., 61:806, 1988.                                 


\bibitem{Rock92}                                                                
S.~Rock {\em et al.}, Phys. Rev. D, 46:24, 1992.                                      


\end{thebibliography}

\end{document}